\pdfoutput=1
\RequirePackage{ifpdf}
\ifpdf % We are running pdfTeX in pdf mode
\documentclass[pdftex]{sigma}
\else
\documentclass{sigma}
\fi

\numberwithin{equation}{section}

\DeclareMathOperator{\diag}{diag}
\DeclareMathOperator{\id}{id}
\begin{document}

\newcommand{\arXivNumber}{1403.1857}

%\allowdisplaybreaks

\renewcommand{\thefootnote}{$\star$}

\renewcommand{\PaperNumber}{063}

\FirstPageHeading

\ShortArticleName{Gauge Theory on Twisted~$\kappa$-Minkowski: Old Problems and Possible Solutions}

\ArticleName{Gauge Theory on Twisted~$\boldsymbol{\kappa}$-Minkowski:\\
Old Problems and Possible Solutions\footnote{This paper is a~contribution to the Special Issue on Deformations of
Space-Time and its Symmetries.
The full collection is available at \href{http://www.emis.de/journals/SIGMA/space-time.html}
{http://www.emis.de/journals/SIGMA/space-time.html}}}

\Author{Marija DIMITRIJEVI\'C~$^\dag$, Larisa JONKE~$^\ddag$ and Anna PACHO{\L} $^{\S}$}

\AuthorNameForHeading{M.~Dimitrijevi\'c, L.~Jonke and A.~Pacho{\l}}

\Address{$^\dag$~University of Belgrade, Faculty of Physics, Studentski trg 12, 11000 Beograd, Serbia}
\EmailD{\href{mailto:dmarija@ipb.ac.rs}{dmarija@ipb.ac.rs}}

\Address{$^\ddag$~Division of Theoretical Physics, Rudjer Bo\v skovi\' c Institute, Bijeni\v cka 54, 10000 Zagreb,
Croatia}
\EmailD{\href{mailto:larisa@irb.hr}{larisa@irb.hr}}

\Address{$^\S$~Science Institute, University of Iceland, Dunhaga 3, 107 Reykjavik, Iceland}
\EmailD{\href{mailto:pachol@hi.is}{pachol@hi.is}}

\ArticleDates{Received March 10, 2014, in f\/inal form June 05, 2014; Published online June 14, 2014}

\Abstract{We review the application of twist deformation formalism and the construction of noncommutative gauge theory
on~$\kappa$-Minkowski space-time.
We compare two dif\/ferent types of twists: the Abelian and the Jordanian one.
In each case we provide the twisted dif\/ferential calculus and consider ${U}(1)$ gauge theory.
Dif\/ferent methods of obtaining a~gauge invariant action and related problems are thoroughly discussed.}

\Keywords{$\kappa$-Minkowski; twist; ${U}(1)$ gauge theory; Hodge dual}

\Classification{81T75; 58B22}

\renewcommand{\thefootnote}{\arabic{footnote}} \setcounter{footnote}{0}

\vspace{-2mm}

\section{Introduction}

Noncommutative (NC) deformations of space-time, introduced either through ef\/fective models or as fundamental property of
a~theory, lead to models that can mimic some of the properties expected in a~quantized theory of gravity.
In particular, the combination of general relativity and quantum mechanics suggests that at the Plank scale the standard
picture of space-time as a~smooth manifold breaks down and should be replaced by some kind of fuzzy or foam-like
space-time.
This property can be naturally realized by NC deformations of space-time, with prominent examples being fuzzy
spaces~\cite{mh2,mh1,mh2+} and, more generally, matrix models~\cite{matrix1, matrix2, matrix3}.
Furthermore, it is expected that our understanding and implementation of symmetries should be modif\/ied when describing
physics close to the Planck scale.
Disentangling the geometry from matter content of a~theory at this scale is non-trivial, and mixing of internal/gauge
symmetries and geometry/dif\/feomorphisms occurs.
In construction of a~physical theory on NC space-time one implements this property by introducing noncommutative gauge
transformations~\cite{jssw1, jssw2}.

The construction of NC gauge theory is an important step in understanding the physics on NC space-time.
Indeed, the main concept in quantum f\/ield theory underlying the success of the Standard model is the principle of local
gauge symmetries.
Therefore an important aspect of the construction of the gauge theory on NC spaces is a~consistent implementation of
local gauge symmetry in combination with non-locality introduced implicitly through a~NC deformation of space-time.
A~key ingredient in such construction is the Seiberg--Witten map~\cite{SW}.
This mapping includes a~set of non-local and non-linear f\/ield redef\/initions relating commutative and NC gauge f\/ields and
parameters.
Most importantly, it enables a~consistent def\/inition of NC gauge theories for arbitrary gauge groups.

\looseness=-1
In describing NC space-time, we substitute the concept of manifold with an algebra of functions on manifold.
NC deformations of such algebra correspond to NC deformations of space-time.
There exist powerful methods for studying deformations of an algebra of functions if this algebra carries
a~representation of a~Hopf algebra.
In this case one f\/irst considers a~deformation of the Hopf algebra, and then uses the Hopf algebra action in order to
induce a~deformation of the algebra on functions on manifold.
Deformation via Drinfel'd twists~\cite{Drinfeld} is an example of such procedure.
One of the main advantages of this formalism is the straightforward way to def\/ine a~dif\/ferential calculus, an important
ingredient in the construction of a~(gauge) f\/ield theory.

In this work our primary interest is to examine compatibility of the local gauge principle with the deformation of
algebra of functions on a~specif\/ic example of NC space-time, the~$\kappa$-Minkowski space-time.
This example of noncommutative space is the most analysed example of non-constant deformation with potentially
interesting phenomenological consequences~\cite{something}.
The (non-trivial) commutation relations of coordinates of the four-dimensional~$\kappa$-Minkowski space-time are of the
Lie-algebra type $[x^0,x^j] = {i}{\kappa}^{-1}x^j$, where~$j$ denotes the space directions and the zeroth component
corresponds to the time direction.
One of the important properties of this NC space-time is that there is a~quantum group symmetry acting on it.
It is a~dimensionfull deformation of the global Poincar\'e group, the~$\kappa$-Poincar\'e
group~\cite{LukPrvi1, LukPrvi2}.
The constant~$\kappa$ has dimension of energy and sets a~deformation scale.
The construction of f\/ield theory on this Lie algebraic type of noncommutative spacetime attracted a~lot of interest, but
was mainly concentrated on the scalar f\/ield theories~\cite{ww,AC,Luk,alex,Melj}.
The problem of constructing gauge f\/ield theory on~$\kappa$-Minkowski space-time was addressed in~\cite{mii1, mii2}.

All these results have shown that the construction of f\/ield theories on this space is plagued with ambiguities, mostly
due to the lack of understanding of the symmetries of NC space.
We proposed to resolve some of these ambiguities by using specif\/ic Abelian twist to introduce deformation in the
algebra~\cite{miU1twisted1,miU1twisted2}.
We constructed the action for the noncommutative ${U}(1)$ gauge f\/ields in a~geometric way, as an integral of a~maximal form.
However, we could not maintain the~$\kappa$-Poincar\'e symmetry; the corresponding symmetry of the
twisted~$\kappa$-Minkowski space was the twisted ${igl}(1,3)$ symmetry.
It turned out that this is a~generic situation;~$\kappa$-Minkowski space-time obtained from~$\kappa$-Poincar\'e algebra
cannot be obtained by twisting the usual Poincar\'e algebra of symmetry~\cite{AnnaLast}.
Still, the extensions of Poincar\'e algebra are amenable to twist formulation, and in this work we concentrate on two
particular examples.
The twists with support in extensions of Poincar\'{e} algebra were also considered previously
in~\cite{Ballesteros1,Ballesteros2,Bu,MSSKG}.

In Sections~\ref{Section2} and~\ref{Section3} we review the basics of the twist formalism and the Seiberg--Witten map; two main concepts we use in our analysis.
Then, in Sections~\ref{Section4} and~\ref{Section5} we introduce two dif\/ferent twists (Abelian and Jordanian).
Commutation relations of coordinates that follow from those twists are commutation relations of the~$\kappa$-Minkowski
space-time; the twisted symmetry algebra is not the~$\kappa$-Poincar\'e algebra.
We construct the twisted dif\/ferential calculus for both deformations.
Finally, we discuss the ${U}(1)$ gauge theory obtained from two dif\/ferent twists.
We describe the problems related to the integration and the construction of the Hodge dual in details and give some
possible solutions.
We extend the analysis of the deformed gauge theory based on the Abelian twist~\cite{miU1twisted1,miU1twisted2} and
compare it with the one based on the Jordanian twist.
We explicitly show how the dif\/ference in the underlying twisted symmetry is manifested at the level of gauge theory action.

\section{Twist formalism}\label{Section2}

Within the twist formalism the NC deformations are introduced by twisting the underlying symmetry of the theory and then
consistently applying the consequences of the deformation on the geometry of space-time itself.
The underlying symmetries are described in the Hopf algebra language and the NC spaces (as deformed algebras of
functions) are Hopf module algebras.
The twisted symmetry does not have the usual dynamical signif\/icance and, in particular, there is no Noether procedure
associated with it.
We view this symmetry as a~prescription that allows us to consistently apply deformation in the theory.
The twist deformation equips the algebra of smooth compactly supported functions $A=C_c^\infty(M)$ with the twisted
$\star$-product and can be represented by deformed, $\star$-commutators of noncommutative
coordinates\footnote{Throughout this paper dif\/ferential forms are smooth and compactly supported.}.
Since we are interested in deformations of space-time symmetries, we concentrate on the Lie algebra of vector f\/ields and
its deformations; generalization to any Lie algebra is straightforward.

\looseness=-1
Vector f\/ields provide an inf\/inite dimensional Lie algebra, its universal enveloping algebra includes linear dif\/ferential
operators.
Those act naturally on the algebra of functions (Hopf module algebra) on a~manifold.
We work in four dimensions and use the Lorentzian signature (mostly minus).
Generalization to~$n$ dimensions is easily done.
The Lie algebra of vector f\/ields is denoted by $\Xi $ and its elements are vector f\/ields $\xi $, which can be written in
the coordinate basis as: $\xi =\xi^{\mu}\frac{\partial}{\partial x^{\mu}}=\xi^{\mu}\partial_{\mu}$.
This algebra generates the dif\/feomorphism symmetry; one can also consider subgroups of $\Xi $ like Poincar\'{e} algebra
or conformal algebra as symmetry groups.
The universal enveloping algebra of $\Xi $ we denote by $U\Xi $.
It can be equipped with the Hopf algebra structure:
\begin{gather}
\lbrack \xi,\eta]= (\xi^{\mu}\partial_{\mu}\eta^{\rho}-\eta^{\mu}\partial_{\mu}\xi^{\rho})\partial_{\rho},
\nonumber\\
\Delta (\xi)= \xi \otimes 1+1\otimes \xi,
\nonumber
\\
\varepsilon (\xi)= 0,
\qquad
S(\xi)=-\xi.
\label{Uxi}
\end{gather}
The f\/irst line is the algebra relation: commutator of two vector f\/ields is a~vector f\/ield.
In the se\-cond line the coproduct of the generator $\xi $ is given; note that it is primitive.
It encodes the Leibniz rule and specif\/ies how the symmetry transformation acts on products of f\/ields/representations.
In the last line, the counit and the antipode maps are given.

A well def\/ined way to deform the symmetry Hopf algebra is via twist.
The twist $\mathcal{F}$ is an invertible element of $U\Xi \otimes U\Xi $ satisfying the following properties:
\begin{enumerate}\itemsep=-1pt
\item[1)] the cocycle condition
\begin{gather}
(\mathcal{F}\otimes 1)(\Delta\otimes\id)\mathcal{F}=(1\otimes \mathcal{F})(\id\otimes \Delta)\mathcal{F},
\label{Twcond1}
\end{gather}

\item[2)] normalization
\begin{gather}
(\id\otimes \epsilon)\mathcal{F} = (\epsilon\otimes\id)\mathcal{F}=1\otimes 1,
\label{Twcond2}
\end{gather}

\item[3)] perturbative expansion
\begin{gather}
\mathcal{F} = 1\otimes 1 + \mathcal{O}(\lambda),
\label{Twcond3}
\end{gather}
\end{enumerate}
where $\lambda $ is a~deformation parameter.
The last property provides an undeformed case at the zeroth order in parameter~$\lambda$.
We shall frequently use the notation (sum over $\alpha =1,2,\dots,\infty $ is understood)
\begin{gather}
\mathcal{F}=\mathrm{f}^{\alpha}\otimes \mathrm{f}_{\alpha},
\qquad
\mathcal{F}^{-1}=\bar{\mathrm{f}}^{\alpha}\otimes \bar{\mathrm{f}}_{\alpha},
\label{Fff}
\end{gather}
where, for each value of $\alpha $, $\bar{\mathrm{f}}^{\alpha}$ and $\bar{\mathrm{f}}_{\alpha}$ are two distinct
elements of $U\Xi $ (and similarly $ \mathrm{f}^{\alpha}$ and $\mathrm{f}_{\alpha}$ are in $U\Xi $)\footnote{Strictly
speaking a~twisted deformation of the Lie algebra $\Xi $ requires a~topological extension of the corresponding
enveloping algebra $U\Xi $ into an algebra of formal power series $U\Xi \lbrack \lbrack \lambda]]$ in the formal
parameter $\lambda $.
Then the twisting element, can be rewritten as a~power series expansion $ \mathcal{F}=1\otimes 1+
\sum\limits_{\alpha=1}^\infty \lambda^\alpha\mathrm{f}^{\alpha}\otimes\mathrm{f}_{\alpha}$.}.
The twist acts on the symmetry Hopf algebra and gives the twisted symmetry (as deformed Hopf algebra)
\begin{gather}
\lbrack \xi,\eta]= (\xi^{\mu}\partial_{\mu}\eta^{\rho}-\eta^{\mu}\partial_{\mu}\xi^{\rho})\partial_{\rho},
\nonumber
\\
\Delta^{\mathcal{F}}(\xi)= \mathcal{F}\Delta (\xi)\mathcal{F}^{-1}
\nonumber
\\
\varepsilon (t^{a})= 0,
\qquad
S^{\mathcal{F}}(\xi)=\mathrm{f}^{\alpha}S(\mathrm{f}_{\alpha})S(\xi)S(\bar{\mathrm{f}}^{\beta})\bar{\mathrm{f}}_{\beta}.
\label{TwistedUg}
\end{gather}
The algebra remains the same, while in general the comultiplication (coproduct) and antipode change.
The whole deformation depends on formal parameters which control classical limit.
Twisted (deformed) comultiplication leads to the deformed Leibniz rule for the symmetry transformations when acting on
product of f\/ields.

Now we use the twist to deform the commutative geometry of space-time (functions, vector f\/ields, exterior algebra of
forms, tensor f\/ields).
As usual in physics, we will work with smooth complex tensor f\/ields.
In particular, we work with smooth functions, smooth exterior algebra of forms, smooth vector f\/ields.
For more details see for example Chapter 1 in~\cite{AlexPhDThesis}.
We consider one particular class of twists, the Abelian twists~\cite{Abelian1,Abelian2}
\begin{gather}
\mathcal{F}=e^{-\frac{i}{2}\theta^{CD}X_{C}\otimes X_{D}}.
\label{AbTwist}
\end{gather}
Here $\theta^{CD}$ is a~constant antisymmetric matrix, $C,D=1,\dots, p$, $p\leq 4$ and $X_{C}=X_{C}^{\mu}\partial_{\mu}$
are commuting vector f\/ields, $[X_{C},X_{D}]=0$.
This twist fulf\/ils the requirements~\eqref{Twcond1}--\eqref{Twcond3}.
We write all results for this particular twist, but we mention if and when they are valid in more general cases (more
general twists).

Applying the inverse of the twist~\eqref{AbTwist} to the usual point-wise multiplication of functions, $\mu (f\otimes
g)=f\cdot g$; $f,g\in A$, we obtain the $ \star $-product of functions
\begin{gather*}
f\star g = \mu \mathcal{F}^{-1}(f\otimes g) = \bar{\mathrm{f}}^{\alpha}(f)\bar{\mathrm{f}}_{\alpha}(g).
%\label{FunctionsStar}
\end{gather*}
The action of the twist ($\bar{\mathrm{f}}^{\alpha}$ and $\bar{\mathrm{f}}_{\alpha}$) on the functions~$f$ and~$g$ is
via the Lie derivative.
This $ \star $-product is noncommutative, associative and in the limit $\theta^{CD}\rightarrow 0$ it reduces to the
usual point-wise multiplication; the last property is guaranteed by~\eqref{Twcond3} and the associativity is guaranteed
by~\eqref{Twcond1}.
In this way we obtain the noncommutative algebra of functions $A^{\mathcal{F}}=(A,\star)$, i.e.\ the noncommutative space-time.
The product between functions and 1-forms is def\/ined as
\begin{gather*}
h\star \omega =\bar{\mathrm{f}}^{\alpha}(h)\bar{\mathrm{f}}_{\alpha}(\omega)
\end{gather*}
with \looseness=-1 an arbitrary 1-form $\omega $.
The action of $\bar{\mathrm{f}}_{\alpha} $ on forms is (again) given via the Lie derivative.
We often use the Cartan's formula for the Lie derivative along the vector f\/ield $\xi $ of an arbitrary form $\omega $
\begin{gather*}
l_{\xi}\omega =\mathrm{d}i_{\xi}\omega +i_{\xi}\mathrm{d}\omega.
%\label{CartanF-la}
\end{gather*}
Here $\mathrm{d}$ is the exterior derivative and $i_{\xi}$ is the contraction along the vector f\/ield $\xi $.

Arbitrary forms form an exterior algebra with the wedge product.
The $\star$-wedge product on two arbitrary forms~$\omega$ and $\omega^{\prime}$ is
\begin{gather*}
\omega\wedge_\star\omega^{\prime}= \bar{\mathrm{f}}^\alpha(\omega) \wedge \bar{\mathrm{f}}_\alpha(\omega^{\prime}).
%\label{WedgeStar}
\end{gather*}
The usual (commutative) exterior derivative $\mathrm{d}: A\rightarrow \Omega$ satisf\/ies:
\begin{gather}
\mathrm{d} (f\star g)=\mathrm{d}f\star g + f\star \mathrm{d}g,
\qquad
\mathrm{d}^2=0,
\nonumber
\\
\mathrm{d} f=(\partial_\mu f) \mathrm{d}x^\mu = (\partial^\star_\mu f) \star \mathrm{d}x^\mu.
\label{Differential}
\end{gather}
The f\/irst property if fulf\/illed because the usual exterior derivative commutes with the Lie derivative which enters in
the def\/inition of the $\star $-product.
Therefore, we will use the usual exterior derivative as the noncommutative exterior derivative.
Note that the last line of~\eqref{Differential} gives a~def\/inition of the $\partial^\star_\mu$ derivatives.

All the properties and def\/initions introduced so far are also valid for a~more general twist.
However with the def\/inition of integral one has to be more careful.
The usual integral is cyclic under the $\star $-exterior products of forms
\begin{gather*}
\int \omega_{1}\wedge_{\star}\omega_{2}=(-1)^{d_{1}\cdot d_{2}}\int \omega_{2}\wedge_{\star}\omega_{1},
%\label{IntCycl}
\end{gather*}
where $d=\deg(\omega)$, $d_{1}+d_{2}=4$ provided that $S(\bar{\mathrm{f}}^{\alpha})\bar{\mathrm{f}}_{\alpha}=1$ holds,
for more details see~\cite{paa}.
One can check that this indeed holds for the Abelian twist~\eqref{AbTwist}.

Note that this approach to deformation of dif\/ferential calculus, i.e.~twisted approach dif\/fers from the bicovariant
dif\/ferential calculi formulation.
More specif\/ically, the covariance condition, i.e.~$L\triangleright (fg)=(L_{(1)}\triangleright f) (L_{(2)}\triangleright g)$,
$L\in U\Xi$, $f,g\in A$ is satisf\/ied in both approaches (the NC space-time is a~Hopf module algebra).
However the bicovariance condition (requiring that all $\mathrm{d}x_\mu$, must be simultaneously left and
right-invariant) is not satisf\/ied in the twisted version.
Moreover it has been shown that in the case of $\kappa $-Minkowski space-time the four-dimensional bicovariant
dif\/ferential calculi does not exist, but one can construct a~f\/ive-dimensional one, which is bicovariant~\cite{Sitarz}.
Alternative approaches to dif\/ferential calculus on~$\kappa$-Minkowski space-time were also considered
in~\cite{Melj-K-J1,Melj-K-J2,Melj-K-J3}.

\section{NC gauge theory and the Seiberg--Witten map}\label{Section3}

In this section we describe how to construct a~NC gauge theory on a~deformed space-time obtained from the Abelian twist~\eqref{AbTwist}.
To achieve our goal we use the enveloping algebra approach and the Seiberg--Witten map, as developed in~\cite{jssw1, jssw2}.
To have more general results, we work with an arbitrary Lie group~$G$, with generators $T^a$ and $[T^a,T^b] =
if^{abc}T^c$.
The obtained results are then easy to specify to the case of ${U}(1)$ gauge group.
Noncommutative f\/ields we label with a~``hat'' and commutative without a~``hat''.
Under the inf\/initesimal NC gauge transformations the NC gauge f\/ield\footnote{Note that we can expand the noncommutative
forms in the coordinate basis in two dif\/ferent ways $\omega = \omega_\mu \star \mathrm{d}x^\mu =
\tilde{\omega}_\mu\mathrm{d}x^\mu$.
The dif\/ference will only be in the components of forms.
Depending on the situation, we will use one or the other expansion, but we will be careful not to mix them.}
$\hat{A}=\hat{A}_\mu\star \mathrm{d} x^\mu$ transforms as
\begin{gather}
\delta_\alpha^\star \hat{A} = \mathrm{d}\hat{\Lambda}_\alpha + i[\hat{\Lambda}_\alpha \stackrel{\star}{,} \hat{A}], %\stackrel
\label{ATr}
\end{gather}
with the NC gauge parameter $\hat{\Lambda}_\alpha$.
The NC gauge parameter is a~noncommutative function valued (as we shall see later) in the enveloping algebra of the
gauge group.
As all NC functions, it is represented by a~function of the commuting coordinates and it is a~power series expansion in
the deformation parameter.
The index~$\alpha$ signals that in the zeroth order of the deformation parameter the NC gauge parameter
$\hat{\Lambda}_\alpha$ reduces to the commutative, Lie algebra-valued gauge parameter $\alpha= \alpha^aT^a$.
Note that this index~$\alpha$ is not related with the index~$\alpha$ in Section~\ref{Section2}, equation~\eqref{Fff}.
We demand that the consistency condition is satisf\/ied, i.e.~transformations~\eqref{ATr} have to close the algebra
\begin{gather}
[\delta_{\alpha}^\star\stackrel{\star}{,}\delta_{\beta}^\star]=\delta_{-i[\alpha, \beta]}^ \star.
\label{NCConsCond}
\end{gather}
This will be the case provided that the gauge parameter $\hat{\Lambda}_\alpha $ is in the enveloping algebra of the
algebra~$g$ of the gauge group~$G$.\footnote{Note that in~\eqref{ATr} $\star$-commutators appear.
These commutators do not close in the Lie algebra, namely having $A=A^aT^a$ and $B=B^aT^a$ leads to
\begin{gather*}
[A\stackrel{\star}{,} B] = \frac{1}{2}(A^a\star B^b + B^b\star A^a)[T^a,T^b] + \frac{1}{2}(A^a\star B^b - B^b\star A^a)\{T^a,T^b\}. %\stackrel
\end{gather*}
Only in the case of ${U}(n)$ in the def\/ining representation the anticommutator of generators is still in the corresponding
Lie algebra.}
However, an enveloping algebra is inf\/inite dimensional and the resulting theory seems to have inf\/initely many degrees of
freedom.
This problem is solved by the Seiberg--Witten map.
The idea of the Seiberg--Witten map is that all noncommutative variables (gauge parameter, f\/ields) can be expressed in
terms of the corresponding commutative variables and their derivatives; then the NC gauge transformations are induced by
the corresponding commutative gauge transformations
\begin{gather}
{\hat A}(A) + \delta_\alpha^\star{\hat A} (A) = {\hat A} (A + \delta_\alpha A),
\label{DeltaA}
\end{gather}
with the commutative gauge f\/ield $A=A^aT^a$ and the commutative gauge parameter $\alpha=\alpha^a T^a$.
In addition, we assume that we can expand all NC variables as power series in noncommutativity parameter $\theta^{CD}$
introduced by the twist.

In the case of NC gauge parameter the expansion is
\begin{gather*}
\hat{\Lambda}_\alpha = \Lambda^{(0)}_\alpha + \Lambda^{(1)}_\alpha + \Lambda^{(2)}_\alpha+\cdots,
\nonumber
\end{gather*}
with $\Lambda^{(0)}_\alpha =\alpha$.
Inserting this expansion into (\ref{NCConsCond}) and expanding all $\star$-products gives a~variational equation for the
gauge parameter $\hat\Lambda_\alpha$.
This equation can be solved to all orders of the deformation parameter.
The zeroth order solution is the commutative gauge parameter~$\alpha$.
The recursive relation between the~$n$th and the $(n+1)$st order solution is given by~\cite{kayhan, PC11}
\begin{gather*}
{\hat{\Lambda}}^{(n+1)}_\alpha = -\frac{1}{4(n+1)}\theta^{CD}\big\{\hat{A}_C \stackrel{\star}{,} l_D {\hat \Lambda}_\alpha\big\}^{(n)},
%\label{RecRelLambda}
\end{gather*}
where $(A\star B)^{(n)} = A^{(n)}B^{(0)} + A^{(n-1)}B^{(1)} + \cdots + A^{(0)}\star^{(1)} B^{(n-1)} + A^{(1)}\star^{(1)}
B^{(n-2)} +\cdots$ includes all possible terms of order~$n$.
We introduced the following notation: ${\hat A}_C = i_{X_C} {\hat A}$ is a~contraction of the 1-form $ \hat{A}$ along
the vector f\/ield $X_C$ and $l_D$ is a~Lie derivative along the vector f\/ield $X_D$.

Solving the equation~\eqref{DeltaA} order by order in the NC parameter the NC gauge f\/ield $\hat{A}$ is expressed in
terms of the commutative gauge f\/ield~$A$.
The recursive solution in this case is given~by
\begin{gather}
\hat{A}^{(n+1)} = -\frac{1}{4(n+1)}\theta^{CD}\big\{\hat{A}_C \stackrel{\star}{,} l_D \hat{A} + \hat{F}_D\big\}^{(n)}, %\stackrel
\label{RecRelA}
\end{gather}
where $L_D \hat{A}= l_D \hat{A} -i[\hat{A}_D \stackrel{\star}{,} \hat{A}]$ and $\hat{F}_D= i_{X_D} \hat{F}$. %\stackrel

Finally, the f\/ield-strength tensor is def\/ined as $\hat{F} = \mathrm{d}{\hat A} -i\hat{A}\wedge_\star \hat{A}$ and it
transforms covariantly under inf\/initesimal NC gauge transformations,
\begin{gather*}
\delta_\alpha^\star \hat{F} = i [\hat{\Lambda} \stackrel{\star}{,} \hat{F}]. %\stackrel
%\label{FTr}
\end{gather*}
The recursive relation for the SW map solution is given by
\begin{gather}
\hat{F}^{(n+1)} = -\frac{1}{4(n+1)}\theta^{CD}\Big(\big\{\hat{A}_C \stackrel{\star}{,} (l_D + L_D) \hat{F}\big\}^{(n)}  %\stackrel %\stackrel...
-[\hat{F}_C \stackrel{\star}{,} \hat{F}_D]^{(n)}\Big),
\label{RecRelF}
\end{gather}
with the 1-form $\hat{F}_C= i_{X_C} \hat{F}$ and the 2-form $L_C \hat{F} = l_C \hat{F} -i[\hat{A}_C \stackrel{\star}{,}\hat{F}]$.
Also, $[\hat{F}_C \stackrel{\star}{,} \hat{F}_D] = \hat{F}_C\wedge_\star \hat{F}_D - \hat{F}_D\wedge_\star \hat{F}_C$.

\begin{remark}\label{remark1}
 The above SW map solutions are written in the language of forms and with the use of the recursive  % \text ?
relations.
One can also expand these relations in orders of the deformation parameter $\theta^{CD}$ and write the solutions for the
components.
These will depend on the particular form of the twist as we will see later on.
In Sections~\ref{Section4} and~\ref{Section5}, we discuss particular examples of the twisted~$\kappa$-Minkowski.
There we will write the component expansions and we will write them up to f\/irst order in the NC parameter.
\end{remark}

\begin{remark}\label{remark2}
The recursive solutions are valid for the Abelian twist.
For a~more general twist one has to solve the SW map order by order in parameter expansion.
\end{remark}

\section{Kappa-Minkowski from an Abelian twist}\label{Section4}

The main object of this review is the $\kappa $-Minkowski space-time.
We will discuss two dif\/ferent ways of twisting that result in $ \kappa $-Minkowski space-time.
The starting point in both approaches is the symmetry algebra of the four dimensional Minkowski space-time, the Poincar\'{e} algebra ${iso}(1,3)$.
It has $10$ generators: $4$ generators of translations $P_{\mu}$ and 6 generators of Lorentz rotations $M_{\mu \nu}$.
The algebra relations are\footnote{We are working with anti-hermitean generators.}
\begin{gather}
\lbrack P_{\mu},P_{\nu}]= 0,
\qquad
\lbrack M_{\mu \nu},P_{\rho}]=\eta_{\nu \rho}P_{\mu}-\eta_{\mu \rho}P_{\nu},
\nonumber
\\
\lbrack M_{\mu \nu},M_{\rho \sigma}]= \eta_{\mu \sigma}M_{\nu \rho}+\eta_{\nu \rho}M_{\mu \sigma}-\eta_{\mu
\rho}M_{\nu \sigma}-\eta_{\nu \sigma}M_{\mu \rho},
\label{PoincareAlg}
\end{gather}
with $\eta_{\mu \nu}=\diag(+1,-1,-1,-1)$.
The universal enveloping algebra of this algebra we label with ${Uiso}(1,3)$.
Besides the algebra relations (\ref{PoincareAlg}) ${Uiso}(1,3)$ can be equipped with the additional structure
\begin{gather*}
\Delta P_{\mu}=P_{\mu}\otimes 1+1\otimes P_{\mu},
\qquad
\varepsilon (P_{\mu})=0,
\qquad
S(P_{\mu})=-P_{\mu}
\nonumber
\\
\Delta M_{\mu \nu}=M_{\mu \nu}\otimes 1+1\otimes M_{\mu \nu},
\qquad
\varepsilon (M_{\mu \nu})=0,
\qquad
S(M_{\mu \nu})=-M_{\mu \nu}.
%\label{hopfPoincare}
\end{gather*}
It is the Hopf algebra we want to deform via twist.
Unfortunately, we cannot choose a~twist from ${Uiso}(1,3)\otimes {Uiso}(1,3)$ and obtain the $ \kappa $-Minkowski space-time
in the same time~\cite{AnnaLast}.
It follows from the fact that the~$\kappa$-deformation of Poincar\'{e} algebra is characterized by a~classical r-matrix
which satisf\/ies inhomogeneous Yang--Baxter equation and one can not obtain the $\kappa $-Poincar\'{e} Hopf algebra and
$\kappa $-Minkowski as its module from an internal twist.
Therefore, in order to obtain the $\kappa $-Minkowski space-time by twisting, we have to enlarge the starting symmetry
algebra.

In our f\/irst example we choose an Abelian twist given by
\begin{gather}
\mathcal{F}=e^{-\frac{i}{2}\theta^{CD}X_C\otimes X_D}
=e^{-\frac{ia}{2} (\partial_0\otimes x^j\partial_j-x^j\partial_j\otimes \partial_0)},
\label{KappaTwist}
\end{gather}
with two commuting vector f\/ields $X_1=\partial_0$ and $X_2=x^j\partial_j$ and indices $j=1,2,3$.
The constant matrix $\theta^{CD}$ is def\/ined as
\begin{gather*}
\theta^{CD} =\left(
\begin{matrix}
0 & a
\\
-a & 0
\end{matrix}
  \right).
\end{gather*}
This twist fulf\/ils the conditions~\eqref{Twcond1},~\eqref{Twcond2} and (\ref{Twcond3}) with the small deformation
parameter $\lambda=a$.
Detailed analysis of the consequences of this twist and the construction of the ${U}(1)$ gauge theory was done
in~\cite{miU1twisted1,miU1twisted2}.
Therefore, we skip some details here and describe the main problems and results.

The vector f\/ield $X_{1}=\partial_0$ generates translations along $x^{0}$ and belongs to the Poincar\'{e} algebra
${iso}(1,3)$.
However, the vector f\/ield $X_{2}=x^{j}\partial_{j}$ belongs to the general linear algebra ${gl}(1,3)$.
Therefore, we have to consider the inhomogeneous general linear algebra ${igl}(1,3)$ as our starting point for the
symmetry analysis and the twist~\eqref{KappaTwist} then def\/ines $U_{{igl}(1,3)}^{\cal F}[[a]]$.
The commutation relations of ${igl}(1,3)$ are
\begin{gather}
\lbrack L_{\mu \nu},L_{\rho \sigma}] =\eta_{\nu \rho}L_{\mu \sigma}-\eta_{\mu \sigma}L_{\rho \nu},
%\nonumber
%\\
\qquad
[P_{\mu},P_{\nu}] = 0,
\qquad
\lbrack L_{\mu \nu},P_{\rho}]=-\eta_{\mu \rho}P_{\nu}.
\label{igl1n}
\end{gather}
The action of the twist~\eqref{KappaTwist} on the ${igl}(1,3)$ algebra follows from~\eqref{TwistedUg} and it has been
analysed in detail in~\cite{Pachol}.
The most important results are: the algebra~\eqref{igl1n} remains the same; since $X_{2}=x^{j}\partial_{j}$ does not
commute with the generators $P_{\mu}$ and $L_{\mu \nu}$ the comultiplication and the antipode change.
In this way we obtain the twisted ${igl}(1,3)$ Hopf algebra instead of the $\kappa $-Poincar\'{e} algebra found
in~\cite{LukPrvi1, LukPrvi2}.

The Lie algebra ${igl}(1,3)$ contains Poincar\'{e} algebra $ {iso}(1,3) $ as a~subalgebra where the
Lorentz generators $M_{\mu \nu}$ are def\/ined by: $M_{\mu \nu}=L_{\mu \nu}-L_{\nu \mu}$.
The algebra $ {igl}(1,3) $, as well as its classical subalgebras act on the algebra of functions~$A$ via f\/irst
order dif\/ferential operators, i.e.~vector f\/ields, def\/ined by the natural representation $L_{\mu
\nu}=x_{\mu}\partial_{\nu}$, $P_{\mu}=\partial_{\mu}$.
The inverse of the twist~\eqref{KappaTwist} def\/ines the $\star$-product between functions (f\/ields) on
the~$\kappa$-Minkowski space-time
\begin{gather}
f\star g=\mu \{\mathcal{F}^{-1} f\otimes g\}
%\label{StarDef}
=\mu \big\{e^{\frac{ia}{2} (\partial_0\otimes x^j\partial_j-x^j\partial_j\otimes \partial_0)} f\otimes g\big\}
\nonumber
\\
\phantom{f\star g}
 =f\cdot g + \frac{ia}{2} x^j\big((\partial_0 f) \partial_j g -(\partial_j f) \partial_0 g\big) + \mathcal{O}\big(a^2\big)
\nonumber
\\
\phantom{f\star g}
 =f\cdot g + \frac{i}{2}C^{\rho\sigma}_\lambda x^\lambda (\partial_\rho f)\cdot (\partial_\sigma g) +
\mathcal{O}\big(a^2\big),
\label{StarPrExp}
\end{gather}
with $C^{\rho\sigma}_\lambda =a(\delta^\rho_0\delta^\sigma_\lambda -\delta^\sigma_0\delta^\rho_\lambda)$.
This product is associative, noncommutative and hermitean
\begin{gather*}
\overline{f\star g} = \bar{g} \star \bar{f}.
\end{gather*}
The usual complex conjugation we label with ``bar".
In the zeroth order (\ref{StarPrExp}) reduces to the usual point-wise multiplication.
Calculating the commutation relations between the coordinates we obtain
\begin{gather*}
\big[x^0 \stackrel{\star}{,} x^j\big] = x^0\star x^j - x^j \star x^0 = ia x^j,
\qquad
\big[x^i \stackrel{\star}{,} x^j\big] =0.
%\label{AbxStarComm}
\end{gather*}
These are the commutation relations of the~$\kappa$-Minkowski space-time with $a=\kappa^{-1}$.

\subsection{Twisted dif\/ferential calculus and integration}

We have seen in Section~\ref{Section2} that the NC exterior derivative is the usual exterior derivative with the
properties~\eqref{Differential}.
We now discuss the specif\/ic properties due to the twist~\eqref{KappaTwist}.

The basis 1-forms are $\mathrm{d}x^\mu$.
Knowing that the action of a~vector f\/ield on a~form is given via Lie derivative one can show that
\begin{gather}
X_1 (\mathrm{d}x^\mu)=0,
\qquad
X_2 (\mathrm{d}x^\mu) = \delta^\mu_j \mathrm{d}x^j,
\label{LieDerdx}
\\
\mathrm{d}x^\mu \wedge_\star \mathrm{d}x^\nu=\mathrm{d}x^\mu \wedge \mathrm{d}x^\nu = - \mathrm{d}x^\nu \wedge
\mathrm{d}x^\mu = -\mathrm{d} x^\nu \wedge_\star \mathrm{d}x^\mu,
\nonumber
\\
f\star \mathrm{d}x^0=\mathrm{d}x^0 \star f,
\qquad
f\star \mathrm{d}x^j = \mathrm{d}x^j \star e^{ia\partial_0}f.
\label{fstardx}
\end{gather}
Since basis 1-forms anticommute the volume form remains undeformed
\begin{gather*}
\mathrm{d}^{4}_\star x:= \mathrm{d}x^0\wedge_\star\mathrm{d}x^1\wedge_\star \cdots \wedge_\star \mathrm{d}x^{3} =
\mathrm{d}x^0\wedge\mathrm{d}x^1\wedge\cdots \wedge \mathrm{d}x^{3} = \mathrm{d}^{4} x.
%\label{VolForm}
\end{gather*}
The $\star$-derivatives follow from~\eqref{Differential} and are given by
\begin{gather*}
\partial^\star_0 = \partial_0,
\qquad
\partial^\star_j = e^{-\frac{i}{2} a\partial_0}\partial_j,
\nonumber
\\
\partial^\star_0 (f\star g) = (\partial^\star_0 f)\star g + f\star (\partial^\star_0 g),
\qquad
\partial^\star_j (f\star g) = (\partial^\star_j f)\star e^{-ia\partial_0}g + f\star (\partial^\star_j g).
%\label{ParcLeibniz}
\end{gather*}
Arbitrary forms $\omega_{1}$ and $\omega_{2}$ do not anticommute, the $ \wedge $ product is deformed to the $\star
$-wedge product
\begin{gather*}
\omega_{1}\wedge_{\star}\omega_{2}\neq (-1)^{d_{1}\cdot d_{2}}\omega_{2}\wedge_{\star}\omega_{1},
\end{gather*}
where $d_{1}$ and $d_{2}$ are the degrees of forms.
However, under the integral forms anticommute
\begin{gather}
\int \omega_{1}\wedge_{\star}\omega_{2}=(-1)^{d_{1}\cdot d_{2}}\int \omega_{2}\wedge_{\star}\omega_{1},
\qquad
\text{with}
\qquad
d_{1}+d_{2}=4.
\label{kappaIntCycl}
\end{gather}
This holds because the twist~\eqref{KappaTwist} fulf\/ils the property $ S(\bar{f}^{\alpha})
\bar{f}_{\alpha}=1$.
The property (\ref{kappaIntCycl}) can be generalized to
\begin{gather*}
\int \omega_{1}\wedge_{\star}\dots \wedge_{\star}\omega_{p}=(-1)^{d_{1}\cdot d_{2}\dots \cdot d_{p}}\int
\omega_{p}\wedge_{\star}\omega_{1}\wedge_{\star}\dots \wedge_{\star}\omega_{p-1},\>{\mbox{with}}d_{1}+d_{2}+\dots+d_{p}=4. %\mbox
%\label{kappaIntCycl2}
\end{gather*}
We say that the integral is cyclic.
This property is very important for construction of NC gauge theories.

\subsection[${U}(1)$ gauge theory]{$\boldsymbol{{U}(1)}$ gauge theory}

In order to construct the NC ${U}(1)$ gauge theory we now use the SW map solutions from Section~\ref{Section2}.
We expand the recursive relations up to the f\/irst order in the NC parameter~$a$ and use the particular form of the twist
(\ref{KappaTwist}).
Also, when writing the expanded solutions for the components of forms we have to use~\eqref{LieDerdx}.

Expanding~\eqref{RecRelA} we obtain the components of the gauge f\/ield $\hat{A} = \hat{A}_\mu\star \mathrm{d}x^\mu$:
\begin{gather*}
\hat{A}_\mu = A_\mu - \frac{a}{2}\delta^j_{\mu}\big(i\partial_0 A_j + A_0A_j\big)+ {\frac{1}{2}} C^{\rho\sigma}_\lambda
x^\lambda\big(F_{\rho \mu}A_\sigma - A_\rho \partial_\sigma A_\mu\big),
%\label{SWAmu}
\end{gather*}
with the commutative gauge f\/ield $A_\mu$ The f\/irst order solution of the f\/ield-strength tensor $\hat{F} =
\frac{1}{2}\hat{F}_{\mu\nu}\star \mathrm{d} x^\mu\wedge_\star \mathrm{d}x^\nu$ follows from~\eqref{RecRelF} and is given
by
\begin{gather}
\hat{F}_{0j}=F_{0j} -\frac{ia}{2}\partial_0F_{0 j} -aA_0F_{0j} + C^{\rho\sigma}_\lambda x^\lambda\big(F_{\rho
0}F_{\sigma j} -A_\rho\partial_\sigma F_{0j}\big),
\label{SWF0j}
\\
\hat{F}_{ij} = F_{ij} -ia\partial_0 F_{ij} - 2aA_0F_{ij} + C^{\rho\sigma}_\lambda x^\lambda\big(F_{\rho i}F_{\sigma
j} - A_\rho\partial_\sigma F_{ij}\big).
\label{SWFij}
\end{gather}
Finally, in order to write a~NC ${U}(1)$ gauge invariant action we need a~NC Hodge dual of the f\/ield-strength tensor~$\hat{F}$.
We label it with $*\hat{F} $; it should have the following properties:
\begin{gather}
\delta_\alpha^\star (*\hat{F}) = i[\hat{\Lambda}_\alpha \stackrel{\star}{,} *\hat{F}],
\label{HDProperties2}
\\
\lim_{a\to 0}(*\hat{F})_{\mu\nu} = \frac{1}{2}\epsilon_{\mu\nu\rho \sigma}F^{\rho\sigma},
\label{HDProperties3}
\end{gather}
with $F^{\rho\sigma} = \eta^{\rho\mu}\eta^{\sigma\nu}F_{\mu\nu}$.
We use the f\/lat metric $\eta_{\mu\nu}$ to raise and lower indices.
The natural guess for the components of the NC Hodge dual
\begin{gather}
(*\hat{F})_{\mu\nu} = \frac{1}{2}\epsilon_{\mu\nu\alpha\beta}\hat{F}^{\alpha\beta},
\label{HDGuess}
\end{gather}
gives a~2-form $*\hat{F} = \frac{1}{2}(*\hat{F})_{\mu\nu}\star \mathrm{d} x^\mu\wedge_\star \mathrm{d}x^\nu$ which does
not fulf\/il the property (\ref{HDProperties2}).
Therefore, the action constructed using~\eqref{HDGuess}
\begin{gather*}
S= \frac{1}{2}\int \hat{F}\wedge_\star *\hat{F},
%\label{HDGuessNCS}
\end{gather*}
is not gauge invariant.
The construction of the Hodge dual on NC spaces turns out to be a~problem in general.
In some simple examples, like~$\theta$ -constant deformation, the natural guess~\eqref{HDGuess} works well, but for
other more complicated deformations this is not the case.
There are dif\/ferent ways of solving (or at least going around) this problem and we describe some of them next.

\subsection{Discussion}

In the following, we discuss three dif\/ferent ways of overcoming the problem of def\/ining the NC Hodge dual.

\textbf{Method 1.}
We introduce a~two form $\hat{Z}=\frac{1}{2}(\hat{Z})_{\mu\nu}\star \mathrm{d}x^\mu\wedge_\star \mathrm{d}x^\nu$ as
\begin{gather*}
\hat{Z} = \frac{1}{2}\epsilon_{\mu\nu\alpha\beta} \hat{Z}^{\alpha\beta}\star \mathrm{d}x^{\mu}\wedge_\star
\mathrm{d}x^{\nu}
%\label{*FX}
\end{gather*}
and demand that it fulf\/ils the properties~\eqref{HDProperties2} and (\ref{HDProperties3}).
Note that $\hat{Z}_{\mu\nu} =\epsilon_{\mu\nu\alpha\beta} \hat{Z}^{\alpha\beta}$.
Then, using the f\/irst property, we solve the SW map for this two form.
Since it transforms in the adjoint representation, (\ref{HDProperties2}), the recursive relation is given by
\begin{gather*}
\hat{Z}^{(n+1)} = -\frac{1}{4(n+1)}\theta^{CD}\big(\big\{\hat{A}_C \stackrel{\star}{,} (l_D + L_D) \hat{Z}\big\}\big)^{(n)}.
%\label{RecRelZ}
\end{gather*}
Expanding this relation up to f\/irst order in the NC parameter~$a$ and using the twist~\eqref{KappaTwist} we obtain
\begin{gather}
\hat{Z}^{0j} = F^{0j} -ia\partial_0F^{0 j} -2aA_0F^{0j}
+ C^{\rho\sigma}_\lambda x^\lambda\big(F_{\rho}^{\;0}F_{\sigma}^{\;j} -A_\rho\partial_\sigma F^{0j}\big), % \;\;
\label{SWZ0j}
\\
\hat{Z}^{ij} = F^{ij} -\frac{ia}{2}\partial_0 F^{ij} - aA_0F^{ij}
+ C^{\rho\sigma}_\lambda x^\lambda\big(F_{\rho}^{\;i}F_{\sigma}^{\;j} - A_\rho\partial_\sigma F^{ij}\big). % \;\;
\label{SWZij}
\end{gather}
The NC gauge invariant action can be written as
\begin{gather}
S_1 = \frac{1}{2}\int \hat{F}\wedge_\star \hat{Z}
 = -\frac{1}{4}\int\big\{2\hat{F}_{0j}\star e^{-ia\partial_0}\hat{Z}^{0j} + \hat{F}_{ij}\star
e^{-2ia\partial_0}\hat{Z}^{ij}\big\}\star \mathrm{d}^4 x.
\label{SgComp}
\end{gather}
The terms $e^{-ia\partial_0}\hat{Z}^{0j}$ and $e^{-2ia\partial_0}\hat{Z}^{ij} $ come from $\star$-commuting basis
1-forms with the compo\-nents~$\hat{Z}^{\mu\nu}$.
Inserting the SW map solutions~\eqref{SWF0j},~\eqref{SWFij},~\eqref{SWZ0j} and~\eqref{SWZij} into~\eqref{SgComp} leads
to
\begin{gather}
S_1 = -\frac{1}{4}\int \mathrm{d}^4 x\left\{F_{\mu\nu}F^{\mu\nu} -\frac{1}{2}C^{\rho\sigma}_\lambda x^\lambda
F^{\mu\nu}F_{\mu\nu} F_{\rho\sigma} + 2C^{\rho\sigma}_\lambda x^\lambda F^{\mu\nu} F_{\mu\rho}F_{\nu\sigma}\right\}.
\label{SgExp}
\end{gather}
This action is invariant under the commutative ${U}(1)$; this is guaranteed by the SW map.
One can further study coupling to the matter f\/ields, equations of motion and their solutions.
It seems that this method works f\/ine.
On the other hand, we had to introduce an additional f\/ield $\hat{Z}$ as a~replacement for the NC Hodge dual of the
f\/ield-strength tensor.
This f\/ield is an independent f\/ield, not a~function of $\hat{F}$ as in the case of the Hodge dual.
After the expansion in the commutative f\/ields, we see that there are no new degrees of freedom; SW map takes care of
that.
However, if one discusses the freedom of the SW map~\cite{LutzSWAmb} one f\/inds additional covariant terms that enter the
action~\eqref{SgExp} with arbitrary coef\/f\/icients.
These terms could be f\/ixed by imposing some additional physical requirements.

\textbf{Method 2.}
If we look closely at the problem of def\/inition of the Hodge dual, we see that the problem arises because the basis
1-forms do not $\star$-commute with functions~\eqref{fstardx}.
But we are free to choose a~dif\/ferent basis, because we write the action in a~basis independent form.
Instead of working in the coordinate basis, we now redo calculations in the natural basis (frame in the sense of
Madore~\cite{Mbooks}).
This basis is def\/ined as a~basis in which basis 1-forms $\theta^a$ $\star$-commute with functions, $f\star \theta^a=
\theta^a\star f$.
Its particular form will in general depend on the choice of the twist.
For the twist~\eqref{KappaTwist} it is given by~\cite{alex}
\begin{gather*}
x^\mu=(t=x^0, x, y, z),
\qquad
\mathrm{d}x^\mu = (\mathrm{d}t, \mathrm{d}x, \mathrm{d}y, \mathrm{d}z),
\qquad
\partial_\mu = (\partial_t, \partial_x, \partial_y, \partial_z)
\nonumber
\\
\downarrow
\nonumber
\\
x^a = (t, r, \theta, \varphi),
\qquad
\theta^a=\left(\mathrm{d}t, \frac{\mathrm{d}r}{r}, \mathrm{d}\theta, \mathrm{d}\varphi\right),
\qquad
e_a=(\partial_t, r\partial_r, \partial_\theta, \partial_\varphi).
%\label{nicebasis}
\end{gather*}
The twist~\eqref{KappaTwist} is then rewritten as
\begin{gather*}
\mathcal{F}=e^{-\frac{i}{2}\theta^{ab}X_a\otimes X_b}=e^{-\frac{ia}{2} (\partial_t\otimes r\partial_r-r\partial_r\otimes \partial_t)}
%\label{NiceTwist}
\end{gather*}
with $X_1=\partial_t=e_0$ and $X_2=r\partial_r=e_1$.
The $\star$-product in this basis is
\begin{gather*}
f\star g=f\cdot g +\frac{ia}{2}\big((e_0f)(e_1g) - (e_1f)(e_0g)\big) + \mathcal{O}\big(a^2\big)
\nonumber
\\
\phantom{f\star g}
=f\cdot g +\frac{ia}{2}\big((\partial_t f)(r\partial_r g) - (r\partial_r f)(\partial_t g)\big) +\mathcal{O}\big(a^2\big).
%\label{NiceStar}
\end{gather*}
Note that the new basis is not f\/lat, the metric is given by $g_{ab} = \diag(1, -r^2, -r^2, -r^2\sin^2\theta)$.
Since the metric does not depend on the coordinate~$t$ the $\star$-inverse is the same as the usual inverse $g_{ab}\star
g^{ac} = g_{ab} g^{ac} = \delta_a^c$.
The volume element is
\begin{gather*}
d^4x= \sqrt{-g} \epsilon_{abcd}\theta^a\wedge\theta^b\wedge\theta^c \wedge\theta^d =
r^2\sin\theta\mathrm{d}t\mathrm{d}r\mathrm{d}\theta\mathrm{d} \varphi.
%\label{NiceVol}
\end{gather*}
The Hodge dual $*\hat{F}$ we def\/ine generalizing the usual expression for the Hodge dual in curved space given by
\begin{gather*}
*F^{(0)} = \frac{1}{2}\epsilon_{abcd}\sqrt{-g}g^{ae}g^{bf}F^{(0)}_{ef} \theta^c\wedge\theta^d.
\end{gather*}
In order to have a~NC gauge invariant action it is necessary that $*\hat{F}$ transforms covariantly under NC gauge
transformations.
To ensure this we have to covariantize the expression $\sqrt{-g}g^{ae}g^{bf}$.
We def\/ine
\begin{gather*}
*\hat{F} = \frac{1}{2}\epsilon_{abcd}\hat{G}^{aebf}\star \hat{F}_{ef}\star \theta^c\wedge_\star\theta^d.
%\label{Nice*F}
\end{gather*}
Here $\hat{G}^{aebf}$ is the quantity that under NC gauge transformations transforms covariantly
\begin{gather*}
\delta^\star_\alpha \hat{G}^{aebf} = i\big[\hat{\Lambda}_\alpha \stackrel{\star}{,} \hat{G}^{aebf}\big].
%\label{NiceTrLawG}
\end{gather*}
and in the limit $a\to 0$ reduces to $\sqrt{-g}g^{ae}g^{bf}$.
The SW map solution for $\hat{G}^{aebf}$ up to f\/irst order in~$a$ is given by
\begin{gather*}
\hat{G}^{aebf} = \sqrt{-g}g^{ae}g^{bf} - aA_0e_1(\sqrt{-g}g^{ae}g^{bf}).
%\label{NiceSWG}
\end{gather*}
The SW map solutions for~$A$ and~$F$ in the new basis are
\begin{gather*}
\hat{A}_a=A_a^0 + \frac{a}{2}\big(A_1^0F^0_{0a} -A_0^0F^0_{1a} + A_1^0(e_0A_a) - A_0^0(e_1A_a)\big),
%\label{NiceSWA}
\\
\hat{F}_{ab}=F^0_{ab} +a\big(F^0_{0a}F^0_{1b} - F^0_{1a}F^0_{0b} -A^0_0(e_1 F^0_{ab}) +A^0_1(e_0 F^0_{ab})\big).
%\label{NiceSWF}
\end{gather*}
Finally, we construct and expand the NC ${U}(1)$ gauge invariant action and obtain
\begin{gather}
S_2=\frac{1}{2}\int (*\hat{F})\wedge_\star \hat{F}
=-\frac{1}{4}\int \mathrm{d}^4 x\big \{F^0_{ab}F^{0ab} + aF^{0ab}\big(4F^0_{0a}F^0_{1b} - F^0_{01}F^0_{ab}\big)\big\},
\label{NiceActionExp}
\end{gather}
with $\mathrm{d}^4 x = r^2\sin\theta\mathrm{d}t\mathrm{d}r\mathrm{d}\theta \mathrm{d} \varphi$.
Note that the action~\eqref{NiceActionExp} has the same form as the expanded action for the NC gauge f\/ield in the case
of~$\theta$-constant deformation~\cite{jssw1, jssw2}.
This is the consequence of the particular choice of basis, the natural basis, in which the twist looks like the
Moyal--Weyl twist.
One can do a~coordinate transformation and write back the action~\eqref{NiceActionExp} in the coordinate basis.
The result (as expected) is~\eqref{SgExp}.
Concerning the degrees of freedom, the situation is the same as in Method~1.
There we introduced the f\/ield $\hat{Z}$, while in Method~2 we introduced the f\/ield $\hat{G}^{aebf}$.
Therefore, the SW map freedom will contribute here as well and (we expect) with the same number of terms, since both
$\hat{Z}$ and $\hat{G}$ transform covariantly.

\textbf{Method 3.}
In~\cite{PLGaugeGr} yet another approach is discussed.
The action for the commutative ${U}(1)$ gauge theory coupled to gravity can be def\/ined as\footnote{In the
paper~\cite{PLGaugeGr} the authors consider a~more general case: non-Abelian gauge theory coupled to gravity.
Here we concentrate only on the ${U}(1)$ theory in f\/lat space-time.}
\begin{gather}
S= \frac{1}{2}\int \epsilon_{\mu\nu\rho\sigma}\left(f^{\mu\nu}F -\frac{1}{12}f_{\alpha\beta}f^{\alpha\beta}V^\mu\wedge V^\nu\right)
\wedge V^\rho\wedge V^\sigma.
\label{PLActionComm}
\end{gather}
Three f\/ields appear in this action: the 2-form f\/ield-strength tensor $F=\frac{1}{2} F_{\mu\nu}\mathrm{d}x^\mu\wedge
\mathrm{d}x^\nu$, the 1-form vierbein $V^\mu=\mathrm{d}x^\mu$ and the auxiliary f\/ield $f_{\alpha\beta}$.
We work in the f\/lat space-time: indices $\mu, \nu,\dots$ are f\/lat % $\mu, \nu,\dots$
indices\footnote{The standard notation in gravity is
that greek indices are curved (Einstein) indices while Latin indices are f\/lat (Lorentz) indices.
However, in our case the greek indices are f\/lat because we work in the f\/lat space-time.
That is, we apply the method of~\cite{PLGaugeGr} but in a~very simple case: f\/lat space-time and ${U}(1)$ gauge theory.}
and are raised and lowered with the f\/lat metric $\eta_{\mu\nu}$.
Under the commutative ${U}(1)$ gauge transformation all three f\/ields are invariant and therefore the
action~\eqref{PLActionComm} is also invariant.
The equation of motion for the f\/ield $f_{\alpha\beta}$ identif\/ies $f_{\alpha\beta}=\frac{1}{2}F_{\alpha\beta}$.
Inserting this into the action~\eqref{PLActionComm} gives
\begin{gather*}
S = -\frac{1}{4}\int \mathrm{d}^4x F_{\mu\nu}F^{\mu\nu}.
%\label{PLActionCommOnshell}
\end{gather*}
We see that by introducing the auxiliary f\/ield $f_{\alpha\beta}$ one can avoid the explicit use of the Hodge dual in the
action.
All this should now be generalized to the NC spaces.

We def\/ine the NC ${U}(1)$ gauge f\/ield action as
\begin{gather}
S_3= \frac{1}{2}\int \epsilon_{\mu\nu\rho\sigma}\left(\frac{1}{2}\big(\hat{f}^{\mu\nu}\star \hat{F}+\hat{F}\star\hat{f}^{\mu\nu}\big)
-\frac{1}{12}\hat{f}_{\alpha\beta}\star \hat{f}^{\alpha\beta}\star \hat{V}^\mu\wedge_\star\hat{V}^\nu\right)
\wedge_\star \hat{V}^\rho\wedge_\star \hat{V}^\sigma,
\label{PLActionNC}
\end{gather}
with noncommutative f\/ields $\hat{f}_{\alpha\beta}$, $\hat{F}$ and $\hat{V}^\mu$.
Note that in order to have a~hermitean action we had to symmetrize the f\/irst term.
All these f\/ields transform covariantly, that is
\begin{gather*}
\delta^\star_\alpha \hat{F} = i\big[\hat{\Lambda}_\alpha \stackrel{\star}{,} \hat{F}\big],
\qquad
\delta^\star_\alpha \hat{f}_{\alpha\beta} = i\big[\hat{\Lambda}_\alpha \stackrel{\star}{,} \hat{f}_{\alpha\beta}\big],
\qquad
\delta^\star_\alpha \hat{V}^\mu = i\big[\hat{\Lambda}_\alpha \stackrel{\star}{,} \hat{V}^\mu\big].
%\label{PLNCTrLaws}
\end{gather*}
The next step is to solve the SW map for these f\/ields and expand the action.
Calculating the equations of motion and inserting them into the expanded action results in the on-shell action.

Solution for the f\/ield-strength tensor we already have; it is given by~\eqref{RecRelF}.
The solutions for other f\/ields are easy to f\/ind since these f\/ields transform in the adjoint representation of the NC~${U}(1)$.
The solutions are given by
\begin{gather*}
\hat{f}_{\alpha\beta}^{(n+1)}=-\frac{1}{4(n+1)}\theta^{CD}\big(\big\{\hat{A}_C \stackrel{\star}{,} (l_D + L_D)\hat{f}_{\alpha\beta}\big\}\big)^{(n)},
%\label{RecRelf}
\\
\hat{V}^{\mu(n+1)}=-\frac{1}{4(n+1)}\theta^{CD}\big(\big\{\hat{A}_C \stackrel{\star}{,} (l_D + L_D)\hat{V}^\mu\big\}\big)^{(n)}.
%\label{RecRelV}
\end{gather*}
Inserting these solutions in the action~\eqref{PLActionNC} and expanding up to f\/irst order in the deformation
parameter~$a$ we obtain
\begin{gather}
S_3=\frac{1}{2}\int \epsilon_{\mu\nu\rho\sigma}\big(f^{\mu\nu}F
-\frac{1}{12}f_{\alpha\beta}f^{\alpha\beta}V^\mu\wedge V^\nu\big)\wedge V^\rho\wedge V^\sigma
\nonumber
\\
\phantom{S_3=}
{}+\frac{1}{4}\theta^{CD}\int \epsilon_{\mu\nu\rho\sigma}\big(F_{CD}f^{\mu\nu}F\wedge V^\rho\wedge V^\sigma +
f^{\mu\nu}F_C\wedge F_D\wedge V^\rho\wedge V^\sigma
\nonumber
\\
\phantom{S_3=}
{}-\frac{1}{12}F_{CD}f_{\alpha\beta}f^{\alpha\beta}V^\mu\wedge V^\nu\wedge V^\rho\wedge V^\sigma\big).
\label{PLActionNCExpanded}
\end{gather}
The one-form $F_C$ is obtained by contraction along the vector f\/ield $X_C$, $F_C = i_{X_C}F$ and the zero-form $F_{CD}$
is def\/ined as a~double contraction $F_{CD} = i_{X_C}i_{X_D} F$ and can be rewritten in terms of Lie derivatives of the
connection one-form~$A$ as $F_{CD}= -l_C A_D + l_D A_C$ with $A_C= i_{X_C}A$.
The vector f\/ields $X_C$, $X_D$ are def\/ined in~\eqref{KappaTwist}.
More explicitly, the expanded action~\eqref{PLActionNCExpanded} is given~by
\begin{gather}
S_3=\int {\rm d}^4x\bigg({-}F^{\alpha\beta}f_{\alpha\beta} + f^{\alpha\beta}f_{\alpha\beta}
-\frac{1}{2}C^{\rho\sigma}_\lambda x^\lambda F_{\rho\sigma}\big({-}F^{\alpha\beta}f_{\alpha\beta}+f^{\alpha\beta}f_{\alpha\beta}\big)
\nonumber
\\
\phantom{S_3=}
{}-C^{\rho\sigma}_\lambda x^\lambda f^{\alpha\beta}F_{\rho\alpha}F_{\sigma\beta}\bigg).
\label{PLActionNCExpanded2}
\end{gather}
Varying this action with respect to $f_{\alpha\beta}$ gives the equation of motion for this f\/ield
\begin{gather*}
f_{\alpha\beta}\left(1 - \frac{1}{2}C^{\rho\sigma}_\lambda x^\lambda F_{\rho\sigma}\right) = \frac{1}{2}F_{\alpha\beta}
-\frac{1}{4}C^{\rho\sigma}_\lambda x^\lambda F_{\rho\sigma}F_{\alpha\beta} +\frac{1}{2}C^{\rho\sigma}_\lambda x^\lambda
F_{\rho\alpha}F_{\sigma\beta},
%\label{PLActionNCEOMf}
\end{gather*}
with the solution up to f\/irst order in the deformation parameter~$a$
\begin{gather*}
f_{\alpha\beta} = \frac{1}{2}F_{\alpha\beta} +\frac{1}{2}C^{\rho\sigma}_\lambda x^\lambda F_{\rho\alpha}F_{\sigma\beta}.
%\label{PLActionNCSolf}
\end{gather*}
Inserting this solution into the action~\eqref{PLActionNCExpanded2} gives to on-shell action
\begin{gather*}
S_3 = -\frac{1}{4}\int \mathrm{d}^4 x\left\{F_{\mu\nu}F^{\mu\nu} -\frac{1}{2}C^{\rho\sigma}_\lambda x^\lambda
F^{\mu\nu}F_{\mu\nu} F_{\rho\sigma} + 2C^{\rho\sigma}_\lambda x^\lambda F^{\mu\nu} F_{\mu\rho}F_{\nu\sigma}\right\} .
%\label{PLActionNCOnShell}
\end{gather*}
This is exactly the result we obtained in Method 1~\eqref{SgExp} and Method 2~\eqref{NiceActionExp}.

Finally, to conclude this analysis: There are dif\/ferent ways to solve (or at least to go around) the problem of the
def\/inition of NC Hodge dual.
What seems to be common for all of them is the introduction of an additional NC f\/ield in the adjoint representation.
These f\/ields do not change the number of degrees of freedom due to the SW map, but they can introduce additional
covariant terms in the expanded actions provided one discusses the freedom of the SW map.
Then these additional terms can be used to render some nice properties of the theory, like
renormalizability~\cite{MajaDusko}.

\section{Kappa-Minkowski from a~Jordanian twist}\label{Section5}

The twisted symmetry of the $\kappa $-Minkowski space-time constructed in Section~\ref{Section4} is the twisted ${igl}(1,3)$.
However, we would like to stay as close as possible to the Poincar\'{e} symmetry, that is we do not want to enlarge the
symmetry algebra too much.
Therefore, in this section we discuss twisting of the Poincar\'{e}--Weyl algebra denoted by ${iwso}(1,3)$.
This algebra has 11 generators: 10 of the Poincar\'{e} algebra and the dilatation generator~$J$.
The algebra is given by~\eqref{PoincareAlg} and additional commutators:
\begin{gather*}
\lbrack M_{\mu \nu},J]=0,
\qquad
\lbrack J,P_{\mu}]=P_{\mu}.
%\label{DilCommRel}
\end{gather*}
In the natural representation the (anti-hermitean) generators are given by $M_{\mu
\nu}=x_{\mu}\partial_{\nu}-x_{\nu}\partial_{\mu}$, $P_{\mu}=\partial_{\mu}$ and $J=-x^{\mu}\partial_{\mu}$.
The universal enveloping algebra of this algebra is ${Uiwso}(1,3)$; it becomes a~Hopf algebra with the
structure~\eqref{Uxi}.

We use the so-called Jordanian twist to deform ${Uiwso}(1,3)$.
Generally Jordanian twists are related with the Borel subalgebra of a~given Lie algebra: $\mathfrak{b}^{2}=\{h,e
\,
|
\,
[h,e] =e\}$.\footnote{$\mathfrak{b}^{2}$ is isomorphic to the 2-dimensional solvable Lie algebra
$\mathfrak{an}^{1}$.}
Such twists have the following form~\cite{Jordanian1,Jordanian2}
\begin{gather*}
\mathcal{F}_{Jor}=\exp \left(h\otimes \sigma \right),
\end{gather*}
where $\sigma =\ln (1+\lambda e)$ with the deformation parameter~$\lambda$.
These kind of twists can be symmetrized as shown in~\cite{Giaquinto,Ohn,Tolstoy}.

In order to have a~hermitean $\star $-product we work with the symmetrized version of Jordanian twist related with the
Borel subalgebra of ${iwso}(1,3)$ given by dilatation~$J$ and momenta $P_0$ generators: $[J,P_0 ]=P_0$.
The inverse of such symmetrized Jordanian twist is given by~\cite{Giaquinto}:
\begin{gather}
\label{FJor}
\mathcal{F}^{-1}=\sum\limits_{m=0}^{\infty}\frac{1}{m!}\left(-\frac{ia}{2}\right)^{m}
\sum\limits_{r=0}^{m}\left(-1\right)^{r}{\binom{m}{r}}P_0^{m-r}J^{\langle r\rangle }\otimes P_0^{r}J^{\langle m-r\rangle },
\end{gather}
where the following notation is used:
\begin{gather*}
J^{\langle 0\rangle }=1,
\qquad
J^{\langle r\rangle } =J(J+1)\cdots (J+r-1),
\qquad
r=1,2,\ldots.
\end{gather*}
Under the action of the twist the algebra relations do not change.
However, the coalgebra sector is deformed.
We give here the deformed coproduct for momenta generators only
\begin{gather}
\label{coP_jor}
\Delta^{\cal F}\left(P_{\mu}\right)
=\sum\limits_{m=0}^{\infty}\left[\left(-1\right)^{m}\left(\frac{ia}{2}\right)^{2m}\left(P_0\otimes
P_0\right)^{m}\Big[\Delta_0\left(P_{\mu}\right) +\frac{ia}{2}\left(P_0\otimes P_{\mu}-P_{\mu}\otimes
P_0\right)\Big]\right].
\end{gather}
It will be used to calculate the coproduct for the new derivatives $\partial_{\mu}^{\star}$ in the next subsection.
For the rest of deformed coproducts we refer the reader to the Appendix.
Once again, the twisted Hopf algebra is not the~$\kappa$-Poincar\'{e} algebra from~\cite{LukPrvi1, LukPrvi2}.
It is the twisted $U_{{iwso}(1,3)}^{\cal F}[[a] ] $.

The inverse of the twist def\/ines the $\star $-product between functions and in a~compact form can be written in the
following way:
\begin{gather*}
f\star g=\mu\big\{\mathcal{F}^{-1}f\otimes g\big\}
\\
\phantom{f\star g}
=\mu \left[\sum\limits_{m=0}^{\infty}
\frac{1}{m!}\left(-\frac{ia}{2}\right)^{m}\sum\limits_{r=0}^{m}\left(-1\right)^{r}{\binom{m}{r}}P_0^{m-r}J^{\langle r\rangle } (f )
\otimes P_0^{r}J^{\langle m-r\rangle } (g )\right].
\end{gather*}
For the future use we rewrite $\mathcal{F}^{-1}$ order by order, using expansion in the deformation parameter~$a$:
\begin{gather*}
\mathcal{F}^{-1}= 1\otimes 1-\frac{ia}{2} (P_0\otimes J-J\otimes P_0 )
\\
\phantom{\mathcal{F}^{-1}=}
{}+\frac{ (ia )^{2}}{8}\left(P_0^{2}\otimes J (J+1 ) -2P_0 J\otimes P_0 J+J (J+1 )
\otimes P_0^{2}\right)
%\label{InvFExpanded}
\\
\phantom{\mathcal{F}^{-1}=}
{}-\frac{ (ia )^{3}}{8}\frac{1}{3!}\big[P_0^{3}\otimes J (J+1 )  (J+2 )
\\
\phantom{\mathcal{F}^{-1}=}
{}-{3}P_0^{2}J\otimes P_0 J (J+1 ) +{3}P_0 J (J+1 ) \otimes P_0^{2}J-J (J+1 )
\left(J+2\right) \otimes P_0^{3}\big]+ \mathcal{O}\big(a^{4}\big).
\end{gather*}
The $\star $-product is then
\begin{gather*}
f\star g= \mu \{\mathcal{F}^{-1}f\otimes g\}
=f\cdot g+i\frac{a}{2}x^{j} (\partial_0 f\partial_{j}g-\partial_{j}f\partial_0 g )
\\
\phantom{f\star g=}
{}+\frac{a^{2}}{8}\big({-}x^{i}x^{j}\big[\partial_0^{2}f (\partial_{i}\partial_{j}g )
+ (\partial_{i}\partial_{j}f ) \partial_0^{2}g-2 (\partial_0\partial_{i}f )
 (\partial_0\partial_{j}g ) \big]
%\label{StarPrExpJordanian}
\\
\phantom{f\star g=}
{}+2\partial_0 f\partial_0 g+2x^{\rho}\partial_0 f\partial_0\partial_{\rho}g+2x^{\mu}\partial_0\partial_{\mu}f\partial_0 g\big)+\mathcal{O}(a^{3}).
\end{gather*}
The commutation relations between coordinates are the $\kappa $-Minkowski commutation relations:
\begin{gather*}
\big[x^{0}\stackrel{\star}{,} x^{j}\big]=x^{0}\star x^{j}-x^{j}\star x^{0}=iax^{j},
\qquad
\big[x^{i}\stackrel{\star}{,} x^{j}\big]=0.
%\label{xStarCommJord}
\end{gather*}

Note that the star product for $\kappa $-Minkowski space-time up to the f\/irst order is the same when coming from the
Jordanian twist (also the non-symmetric one, see, e.g.,~\cite{Pachol}) and from the Abelian twist~\eqref{StarPrExp}.

\subsection{Twisted dif\/ferential calculus}

The usual exterior derivative is the $\star $-exterior derivative.
As in Section~\ref{Section4}, we use the coordinate basis; the basis 1-forms are $\mathrm{d}x^{\mu}$.
As the action of a~vector f\/ield on a~form is given via Lie derivative, we obtain
\begin{gather*}
P_0 (\mathrm{d}x^{\mu})=0,
\qquad
J(\mathrm{d}x^{\mu})=-\mathrm{d}x^{\mu}.
%\label{LieDerdxJordanian}
\end{gather*}
Using these relations one can show that the basis 1-forms anticommute
\begin{gather*}
\mathrm{d}x^{\mu}\wedge_{\star}\mathrm{d}x^{\nu}=\mathrm{d}x^{\mu}\wedge \mathrm{d}x^{\nu}=-\mathrm{d}x^{\nu}\wedge
\mathrm{d}x^{\mu}=- \mathrm{d}x^{\nu}\wedge_{\star}\mathrm{d}x^{\mu},
\end{gather*}
but do not $\star $-commute with functions
\begin{gather}
f\star \mathrm{d}x^{\mu}= f\mathrm{d}x^{\mu}+\frac{ia}{2}\partial_0 f \mathrm{d}x^{\mu},
\qquad
%\nonumber
%\\
\mathrm{d}x^{\mu}\star f= f\mathrm{d}x^{\mu}-\frac{ia}{2}\partial_0 f \mathrm{d}x^{\mu}.
\label{fstardxJordanian}
\end{gather}
Note that the relations~\eqref{fstardxJordanian} are valid to all orders in $ a$.

One can rewrite the usual exterior derivative of a~function using the $\star $-product as
\begin{gather*}
\mathrm{d}f= (\partial_{\mu}f)\mathrm{d}x^{\mu}=(\partial_{\mu}^{\star}f)\star \mathrm{d}x^{\mu},
%\label{ExtDeriv}
\end{gather*}
where the new derivatives $\partial_{\mu}^{\star}$ are def\/ined by this equation.
We obtain the following relation, again valid to all orders in~$a$:
\begin{gather*}
\left(1+\frac{ia}{2}\partial_0\right) \partial_{\mu}^{\star}=\partial_{\mu}.
\end{gather*}
However, the coproduct for the new derivatives $\partial_{\mu}^{\star}$ can only be calculated order by order,
using~\eqref{coP_jor} and the expansion in~$a$.
Up to second order we obtain:
\begin{gather*}
\Delta (\partial_{\mu}^{\star})=\partial_{\mu}^{\star}\otimes 1+1\otimes
\partial_{\mu}^{\star}-ia\partial_{\mu}^{\star}\otimes \partial_0^{\star}+\mathcal{O}\big(a^{3}\big).
%\label{ParcLeibnizJordanian}
\end{gather*}

Unfortunately, the twist~\eqref{FJor} does not fulf\/il $S(\bar{f}^{\alpha}) \bar{f}_{\alpha}= 1$ and the
integral will not be cyclic.
This is a~problem when one wants to discuss NC gauge theories and use the variational principle.
We postpone the discussion until next subsection.

\subsection[${U}(1)$ gauge theory]{$\boldsymbol{{U}(1)}$ gauge theory}

Having def\/ined dif\/ferential calculus we are ready to formulate NC gauge theory on~$\kappa$-Minkowski space-time
possessing the twisted Weyl--Poincar\'{e} symmetry.
We use the SW map solutions from Section~\ref{Section3} and expand everything up to the f\/irst order in the deformation parameter~$a$.
Strictly speaking, formulae from the Section~\ref{Section3} are only valid in the case of Abelian twist deformation and will not be
valid in the case of Jordanian deformation~\eqref{FJor}.
But since the f\/irst order of the $\star$-product is the same for both twists we can use these solutions up to f\/irst
order\footnote{One can also solve the SW map equations from a~scratch, order by order in the case of Jordanian twist.
We did that and found that the f\/irst order solutions coincide with the f\/irst order expansion of solutions in Section~\ref{Section3}.
However, higher order solutions will be dif\/ferent for the Abelian and the Jordanian deformation.}.

Let us write the f\/irst order expansions of the SW map solutions.
The NC gauge parame\-ter~$\hat{\Lambda}_\alpha$ up to f\/irst order in the NC parameter~$a$ is given by
\begin{gather*}
\hat{\Lambda}_{\alpha} = \alpha -\frac{1}{2}C^{\mu\nu}_\lambda x^\lambda A^0_\mu \partial_\nu\alpha.
%\label{SWLambdaJ}
\end{gather*}
The NC gauge f\/ield $\hat{A}= \hat{A}_\mu\star\mathrm{d}x^\mu$ we calculate from~\eqref{RecRelA}
\begin{gather*}
{\hat A}_\mu = A_\mu - \frac{a}{2}\big(i\partial_0 A_\mu + A_0A_\mu\big) - {\frac{1}{2}} C^{\rho\sigma}_\lambda
x^\lambda A_\rho\big(\partial_\sigma A^0_\mu + F_{\sigma \mu}\big).
%\label{SWAmuJ}
\end{gather*}
The components of the f\/ield-strength tensor $\hat{F} = \frac{1}{2}\hat{F}_{\mu\nu}\star \mathrm{d}x^\mu\wedge_\star
\mathrm{d}x^\nu$ follow from~\eqref{RecRelF}
\begin{gather}
\hat{F}_{\mu\nu} = F_{\mu\nu} -ia\partial_0 F_{\mu\nu} - 2aA_0 F_{\mu\nu} - C^{\rho\sigma}_\lambda x^\lambda
\big(A_\rho\partial_\sigma F_{\mu\nu}- F_{\rho \mu}F_{\sigma \nu}\big).
\label{SWFmunuJ}
\end{gather}
Note that this solution is dif\/ferent then~\eqref{SWF0j} and~\eqref{SWFij}.
This is a~consequence of the Jordanian deformation and the dif\/ference in the dif\/ferential calculus,
compare~\eqref{fstardx} and~\eqref{fstardxJordanian}.
Next step is the construction of gauge invariant action.
Here we face the following problems:
\begin{enumerate}\itemsep=0pt
\item In order to write a~NC action for the gauge f\/ields, we need a~NC generalization of the Hodge dual of the
f\/ield-strength tensor $\hat{F}$.
The problem is the same as in Section~\ref{Section4}.

\item The integral is not cyclic.
\end{enumerate}

In Section~\ref{Section4} we saw that there are three ways to ``step around" the problem~1. % Problem ?
In the case of Jordanian deformation, due to the non-cyclicity of the integral, problems 1 and 2 interfere and cannot be
analyzed separately.
To solve this we will use a~modif\/ication of the f\/irst method in Section~\ref{Section4}.
The other methods we comment in the next subsection.

We modify the integral by introducing a~measure function $\mu(x)$ in the following way
\begin{gather}
\int \mu(x)\cdot (\omega_1 \wedge_\star \omega_2) = \int \mu(x)\cdot (\omega_1 \wedge \omega_2) .
\label{CorrInt}
\end{gather}
Of course, to def\/ine an integral $\omega_1 \wedge_\star \omega_2$ has to be a~maximal form.
Now one has to perform the explicit calculation, expanding the $\star$-product and taking into the account that vector
f\/ields from the def\/inition of the twist act on forms via the Lie derivative and that Lie derivatives fulf\/il the Leibniz
rule.
We do the calculation up to the f\/irst order in the deformation parameter, but it can be generalized to higher orders.
The f\/irst order of~\eqref{CorrInt} is given by
\begin{gather*}
-\frac{ia}{2}\int \mu(x)\big((l_{P_0}\omega_1) \wedge (l_J\omega_2) - (l_J\omega_1) \wedge (l_{P_0}\omega_2)\big)
\\
\qquad
=\frac{ia}{2}\int\big((l_{P_0}\mu)\omega_1 \wedge l_J\omega_2 + \mu\omega_1\wedge l_{[P_0, J]}\omega_2 -
(l_J\mu)\omega_1\wedge (l_{P_0}\omega_2)\big) .
\end{gather*}
Going from the f\/irst to the second line we performed integration by parts and discarded the surface terms.
Knowing that $[J, P_0] = P_0$, we obtain the following conditions on the measure function~$\mu$: $l_{P_0} \mu =0$ and $l_J\mu=-\mu$.
We checked that no new conditions on $\mu(x)$ appear in the second order, and we conjecture that this holds to all
orders.

We see that the measure function is~$a$-independent and does not vanish in the limit $a\to 0$.
One possible solution in four dimensions is given by
\begin{gather}
f(x) = \sqrt{(x_1)^2 + (x_3)^2 + (x_3)^2}.
\label{measure}
\end{gather}
A~more precise mathematical description (and justif\/ication) of adding the measure function (changing the volume element)
can be found in~\cite{MathMeasure}.

Next, we need a~Hodge dual to write a~gauge invariant action for the NC gauge f\/ield.
In addition, since the measure function~$\mu$ does not vanish in the commutative limit, we have to f\/ind a~way to cancel
it from the zeroth order of the equations of motion.
Having these two requirements in mind, we construct the following action
\begin{gather}
S = \frac{1}{2}\int \mu(x) \cdot\big(\hat{Y} \wedge_\star \hat{F}\big),
\label{FrAction}
\end{gather}
where $\mu(x)$ is def\/ined by~\eqref{CorrInt} and~\eqref{measure} and $\hat{Y}$ is a~2-form which satisf\/ies
\begin{gather}
\hat{Y}=\frac{1}{2}\hat{Y}_{\mu\nu}\star \mathrm{d}x^\mu\wedge_\star \mathrm{d}x^\nu,
%\label{FrProperties1}
\\
\delta_\alpha^\star \hat{Y}=i[\Lambda_\alpha \stackrel{\star}{,} \hat{Y}],
\label{FrProperties2}
\\
\lim_{a\to 0}\hat{Y}_{\mu\nu}=\frac{1}{\mu(x)}\frac{1}{2} \epsilon_{\mu\nu\rho\sigma}F^{\rho\sigma}.
\label{FrProperties3}
\end{gather}
The action~\eqref{FrAction} is gauge invariant and the good commutative limit is ensured by~\eqref{FrProperties3}.
Expanding the action~\eqref{FrAction} we obtain
\begin{gather}
S = \frac{1}{2}\int \mu(x) \cdot\frac{1}{4}\big(\hat{Y}_{\mu\nu} \star
\big((1-2ia\partial_0)\hat{F}_{\rho\sigma}\big)\star\epsilon^{\mu\nu\rho\sigma}\mathrm{d}^4x\big).
\label{FrActionExp1}
\end{gather}
The SW map solution for $\hat{Y}$ can be found solving perturbatively~\eqref{FrProperties2} and up to f\/irst order in~$a$
is given by
\begin{gather}
\hat{Y}_{\mu\nu} = \frac{1}{2\mu(x)}\epsilon_{\mu\nu\rho\sigma}\big(F^{\rho\sigma} - a\big(i\partial_0 F^{\rho\sigma} +
A_0F^{\rho\sigma} + C^{\alpha\beta}_\lambda x^\lambda A_\alpha \partial_\beta F^{\rho\sigma}\big)\big).
\label{SWmapFr}
\end{gather}
Inserting the SW map solutions~\eqref{SWFmunuJ} and~\eqref{SWmapFr} into the action~\eqref{FrActionExp1} we obtain
\begin{gather}
S = -\frac{1}{4}\int \mathrm{d}^4x\left(F_{\mu\nu}F^{\mu\nu} + C^{\rho\sigma}_\lambda x^\lambda
\left(F^{\mu\nu}F_{\mu\rho}F_{\nu\sigma} -\frac{1}{2} F^{\mu\nu}F_{\mu\nu} F_{\rho\sigma}\right)\right).
\label{FrActionExpFinal}
\end{gather}
It is obvious that the obtained action is invariant under the commutative ${U}(1)$ gauge transformations, that is
guaranteed by the SW map.
The result is dif\/ferent than the one obtained in Section~\ref{Section4}.
We obtained the same terms as in~\eqref{SgExp}, however the relative coef\/f\/icient between the second and the third term
is dif\/ferent compared with~\eqref{SgExp}.
This dif\/ference can be attributed to the properties of Jordanian twist, i.e., to the dif\/ferent twisted symmetry than in
the previous case.

We can write both actions,~\eqref{SgExp} and~\eqref{FrActionExpFinal}, as
\begin{gather*}
S=-\frac{1}{4}\int {\rm d}^4x\left(F_{\mu\nu}F^{\mu\nu}+C^{\rho\sigma}_{\lambda}x^\lambda\left(n
F^{\mu\nu}F_{\mu\rho}F_{\nu\sigma}-\frac{1}{2}F^{\mu\nu}F_{\mu\nu}F_{\rho\sigma} \right)\right),
\end{gather*}
where the parameter~$n$ takes value $n=2$ in the case of the Abelian twist and $n=1$ in the case of the Jordanian twist.
The corresponding equations of motions are:
\begin{gather}
\partial_\mu F^{\mu\nu} = a\frac{(6n-4)}{4}F^{\nu\mu}F_{0\mu}-a\frac{(3-2n)}{4}\delta^\nu_0
F^{\alpha\beta}F_{\alpha\beta}-C^{\rho\sigma}_{\lambda}x^\lambda\frac{n}{2} \left(F^\mu_\rho\partial_\mu
F^\nu_\sigma+F_{\mu\sigma}\partial_\rho F^{\mu\nu}\right)
\nonumber
\\
\phantom{\partial_\mu F^{\mu\nu} =}
{}-\frac{1}{4} C^{\rho\sigma}_{\lambda}x^\lambda(n-2)F^{\mu\nu}\partial_\mu F_{\rho\sigma}
-\frac{1}{4}C^{\nu\sigma}_{\lambda}x^\lambda\left(1-\frac{n}{2}\right)\partial_\sigma\big(F_{\alpha\beta}F^{\alpha\beta}\big).
\label{eom}
\end{gather}
We notice that the last two terms in~\eqref{eom} are non-zero only for $n=1$, i.e., only in the case of the Jordanian
twists.
Moreover, the second term on the right-hand side changes sign in going from one case to the other.
We conclude (on the level of action and equations of motion) that dif\/ferent twists lead to dif\/ferent physics.
Next step would be the analysis of possible
solutions of these equations, and we hope to address this issue in future.

\subsection{Discussion}

Modifying the f\/irst method of Section~\ref{Section4}, we managed to construct the gauge invariant action in the case of the Jordanian twist.
However, the origin of the measure function~$\mu$ is not very clear.
We can speculate that in higher dimensions the twist~\eqref{FJor} looks simpler and when doing the dimensional reduction
to four dimensions the twist obtains its Jordanian form and the measure function appears in the integral.
So far we were not able to f\/ind a~proper higher dimensional theory that could give all this.

There were two other methods discussed in Section~\ref{Section4}.
Let us comment brief\/ly how they do (not) apply here.
It is obvious that they have to be modif\/ied due to the non-cyclicity of the integral.

One can f\/ind a~natural basis in the case of Jordanian deformation and it is given by
\begin{gather*}
x^\mu=(t=x^0, x, y, z),
\qquad
\mathrm{d}x^\mu = (\mathrm{d}t, \mathrm{d}x, \mathrm{d}y, \mathrm{d}z),
\qquad
\partial_\mu = (\partial_t, \partial_x, \partial_y, \partial_z)
\\
\downarrow
\\
x^a = (t, r, \theta, \varphi),
\qquad
\theta^a=\left(\frac{\mathrm{d}t}{r}, \frac{\mathrm{d}r}{r}, \mathrm{d}\theta, \mathrm{d}\varphi\right),
\qquad
e_a=(r\partial_t, r\partial_r, \partial_\theta, \partial_\varphi).
%\label{nicebasisJordanian}
\end{gather*}
The vector f\/ields entering the twist~\eqref{FJor} are rewritten as $X_1=\partial_0 = \frac{1}{r}e_0$ and
$X_2=-x^\mu\partial_\mu =- \frac{t}{r}e_0 - e_1$.
Then one can check that
\begin{gather*}
X_1 (\theta^a) = 0,
\qquad
X_2 (\theta^a) =0.
\end{gather*}
The basis 1-forms $\theta^a$ are frame 1-forms, they $\star$-commute with functions:
\begin{gather*}
\theta^a \star f = f\star \theta^a = f\cdot \theta^a.
%\label{fthetaJordanian}
\end{gather*}
The construction of the Hodge dual is done following the same steps as in Section~\ref{Section4}.
Only when writing the integral, one has to be careful and add the measure function.
Of course, the $a\to 0$ limit of Hodge dual has to be modif\/ied to cancel the measure function, similar
to~\eqref{FrProperties3}.
This basis simplif\/ies calculations but does not lead to a~big improvement.
The measure function in this basis is $\mu(x) =r$.

The method of an auxiliary f\/ield $f^{ab}$ introduced in~\cite{PLGaugeGr} again does not improve a~lot.
In the sense of the ambiguities of the SW map all three methods are the same as the methods discussed in Section~\ref{Section4}.
All this suggests that the question of NC gauge theory on a~Jordanian~$\kappa$-Minkowski needs to be understood better.

\section{Conclusions}

In this paper we demonstrated how one could construct NC gauge theory consistent with deformation of algebra of
functions on~$\kappa$-Minkowski space-time.
Two key ingredients were twist formalism and the Seiberg--Witten map.
We used Drinfel'd twist to deform the Hopf algebra of symmetry generators and then used the Hopf algebra action to induce
deformation of geo\-met\-ry.
This procedure provides dif\/ferential calculus needed for the construction of a~f\/ield theory.
As a~next step, we def\/ined NC gauge transformations.
The Seiberg--Witten map insures that these NC gauge transformations are actually induced by the corresponding
commutative gauge transformation.
Expanding in the deformation parameter led to ef\/fective models which could be seen as a~possible non-local and
non-linear extension of classical electrodynamics.
Moreover, we showed that dif\/ferent underlying symmetries, ${igl}(1,3)$ and ${iwso}(1,3)$, led to two dif\/ferent deformations
of the standard theory.

We also described in details the obstacles encountered in our analyses and of\/fered some possible solutions.
The failure of the Jordanian twist to provide cyclic integral could be understood as an indication that the underlying
symmetry ${iwso}(1,3)$ and its twisted version are not compatible with the f\/lat metric.
The measure we introduced, seemingly ad hoc, might be seen as a~consistency requirement.
The obstruction we have encountered in the construction of the Hodge dual f\/ield-strength tensor is a~manifestation of
the fact that the introduction of a~NC geometrical structure prevents decoupling of dif\/feomorphisms and gauge
symmetries.
The Hodge dual f\/ield-strength tensor includes both gauge and metric degrees of freedom.
Consistent NC deformation of both gauge and geometry imposes that the metric degrees of freedom should transform
covariantly under the gauge transformation.
This brings in mind the ideas of generalized geo\-met\-ry, a~framework in which one organizes and extends the gauge
transformations and the dif\/feomorphisms within ${O}(d,d)$ group.
Applying these ideas in the present context would imply extending our analyses to quasi-Hopf algebras~\cite{quasiHopf1,quasiHopf2,Schupp} and/or Hopf algebroids~\cite{algebroid1,algebroid2, algebroid3}.

\appendix

\section{Appendix}
One can write the symmetrized version of Jordanian twist (which inverse is given in equation~\eqref{FJor}, see
also~\cite{Giaquinto}) by taking the formal expansion in the parameter~$a$ as:
\begin{gather*}
\mathcal{F}=1\otimes 1+\frac{ia}{2} (P_0\otimes J-J\otimes P_0 )
%\label{F2}
\\
\phantom{\mathcal{F}=}
{}+\frac{ (ia )^{2}}{4}\frac{1}{2}\big(P_0^{2}\otimes J^{\underline{2}}-2P_0 J\otimes JP_0 -2JP_0\otimes
P_0 J+2P_0 J\otimes P_0 J+J^{\underline{2}}\otimes P_0^{2}\big)
\\
\phantom{\mathcal{F}=}
{}-\frac{ (ia )^{3}}{8}\bigg\{\frac{1}{6}J^{\underline{3}}\otimes P_0^{3}-\frac{1}{6}P_0^{3}\otimes
J^{\underline{3}}+\frac{1}{2} JP_0^{2}\otimes JP_0 J
\\
\phantom{\mathcal{F}=}
{}-\frac{1}{2}JP_0 J\otimes JP_0^{2}+P_0 J\otimes P_0^{2}-P_0^{2}\otimes P_0 J\bigg\}+\mathcal{O}\big(a^{4}\big),
\end{gather*}
where $J^{\underline{r}} =J(J-1)\cdots (J-r+1)$, $r=1,2,\ldots$.
The deformed coproducts for Lorentz and dilatation generators can be also calculated order by order in the deformation
parameter~$a$ and up to the third order are the following:
\begin{gather*}
\Delta_{{\rm up}(a^{3} )}^{\cal F} (M_{i} )= \Delta_0 (M_{i} ),
\\
\Delta_{{\rm up}(a^{3} )}^{\cal F} (N_{k} )= \Delta_0 (N_{k} ) +\frac{ia}{2} (P_{k}\otimes J-J\otimes P_{k} )
\\
\phantom{\Delta_{{\rm up}(a^{3} )}^{\cal F} (N_{k} )=}
{}-\frac{ (ia )^{2}}{4}(P_{k}\otimes P_0 J+P_0 J\otimes P_{k}+P_{k}P_0\otimes J+J\otimes P_{k}P_0)
%\label{coN}
\\
\phantom{\Delta_{{\rm up}(a^{3} )}^{\cal F} (N_{k} )=}
{}+\frac{ (ia )^{3}}{8}\big\{P_{k}P_0^{2}\otimes J-J\otimes P_{k}P_0^{2}+P_{k}P_0\otimes P_0 J-P_0 J\otimes P_{k}P_0\big\}+\mathcal{O}\big(a^{4}\big),
\\%[2mm]
\Delta_{{\rm up} (a^{3} )}^{\cal F} (J )= \Delta_0 (J ) +\frac{ia}{2} (P_0\otimes
J-J\otimes P_0 )
\\
\phantom{\Delta_{{\rm up} (a^{3} )}^{\cal F} (J )=}
{}-\frac{ (ia )^{2}}{4} (P_0\otimes P_0 J+P_0 J\otimes P_0 +P_0^{2}\otimes J+J\otimes
P_0^{2} )
\\
\phantom{\Delta_{{\rm up} (a^{3} )}^{\cal F} (J )=}
{}+\frac{ (ia )^{3}}{8}\big\{P_0^{3}\otimes J-J\otimes P_0^{3}-P_0 J\otimes P_0^{2}+P_0^{2}\otimes P_0 J\big\}+\mathcal{O} \big(a^{4}\big).
%\label{coD}
\end{gather*}
Here we introduced the following notation for Poincar\'{e} algebra generators $M_{i}=\frac{1}{2}\epsilon_{ijk}M_{jk}$
for rotations and $N_{i}=M_{0i}$ for boosts.
The twisted coproduct for momenta written in a~compact form~\eqref{coP_jor} in Section~\ref{Section5} can be also expanded:
\begin{gather*}
\Delta_{{\rm up}(a^{3} )}^{\cal F} (P_{\mu} )= \Delta_0 (P_{\mu} )
+\frac{ia}{2} (P_0\otimes P_{\mu}-P_{\mu}\otimes P_0 )
-\frac{ (ia )^{2}}{4} (P_0 P_{\mu}\otimes P_0 +P_0\otimes P_0 P_{\mu} )
\\
\phantom{\Delta_{{\rm up}(a^{3})}^{\cal F}(P_{\mu})=}
{}+\frac{(ia)^{3}}{8}\big\{P_0 P_{\mu}\otimes P_0^{2}-P_0^{2}\otimes P_0 P_{\mu}\big\}+\mathcal{O}\big(a^{4}\big).
\end{gather*}
Twisted antipodes are the following:
\begin{gather*}
S^{\cal F} (M_{i} )= -M_{i},
\\
S^{\cal F} (N_{k} )= -N_{k}+\frac{ia}{2}\left(P_{k}J-\left(1+\frac{ia}{2}P_0\right)
J\left(1+\frac{ia}{2}P_0\right)^{-1}P_{k}\right),
\\
S^{\cal F} (J )= -\left(1+\frac{ia}{2}P_0\right) J\left(1+\frac{ia}{2}P_0\right)^{-1},
\\
S^{\cal F} (P_{\mu} )= -P_{\mu}.
\end{gather*}
To complete the def\/inition of $U_{{iwso}(1,3)}^{\cal F}[[a]]$ we mention that counits stay undeformed:
\begin{gather*}
\epsilon  (M_{i} ) =\epsilon  (N_{k} ) =\epsilon  (P_{\mu} ) =\epsilon  (J ) =0.
\end{gather*}

\subsection{Acknowledgements}
The authors are grateful to P.~Aschieri for pointing out the reference~\cite{Giaquinto} and to A.~Borowiec for
discussions.
We would also like to thank the anonymous referees for their insightful comments.
The work is partially supported by ICTP - SEENET-MTP Project PRJ09 ``Cosmology and Strings'' in frame of the
Southeastern European Network in Theoretical and Mathematical Physics.
A.P.~acknowledges the f\/inancial support of Polish NCN Grant 2011/01/B/ST2/03354.
The work of M.D.~is supported by Project No.~171031 of the Serbian Ministry of Education and Science.

\pdfbookmark[1]{References}{ref}
\LastPageEnding

\end{document}